\documentclass[prb,aps,espf,preprint]{revtex4}
\usepackage{amsmath}
\usepackage{graphicx}
\usepackage{color}

\usepackage{subfigure}

\begin{document}

\title {Ground state structure of BaFeO$_{3}$: Density Functional Theory Calculations}

\author{Gul Rahman}\email{gulrahman@qau.edu.pk}
\author{Saad Sarwar}
\address{Department of Physics,
Quaid-i-Azam University, Islamabad 45320, Pakistan}

\begin{abstract}
Using density functional theory calculations, the ground state structure 
of BaFeO$_3$ (BFO) is investigated with local spin density approximation (LSDA). Cubic, tetragonal, orthorhombic, and rhombohedral types BFO are considered to calculate the formation enthalpy.  The formation enthalpies reveal that cubic is the most stable structure of BFO. Small energy difference between the cubic and tetragonal suggests a possible tetragonal BFO. Ferromagnetic(FM) and anitiferromagnetic (AFM) coupling between the Fe atoms show that all the striochmetric BFO are FM. The energy difference between FM and AFM shows room temperature ferromagnetism in cubic BFO in agreement with the experimental work. 
The LSDA calculated electronic structures are metallic in all studied crystallographic phases of BFO. Calculations including the Hubbard potential $U,i.e.$ LSDA+$U$, show that all phases of BFO are half-metallic consistent with the integer magnetic moments. The presence of half-metallicity is discussed in  terms of electronic band structures of BFO.    
\end{abstract}

\maketitle


\section{Introduction}
Perovskite materials are very important both from theory and experiment point of view due to  ferroelectricity,\cite{1} spin dependent transport and magnetic properties. \cite{2}
Magnetic oxides are helpful in understanding the magnetic coupling through nanostructured interfaces.\cite{10,11,12} 
Particularly, those perovskite oxides that exhibit magnetic and ferroelectric characteristics simultaneously, known as multiferroics, can
have practical device applications such as spin transistor memories, whose magnetic properties can be tuned by electric fields through the lattice strain effect.\cite{G1}
Iron based perovskite oxides have very interesting properties due to the different oxidation states of Fe, which gives rise to different crystal structures and stoichiometries.
A small number of oxides containing Fe in high valence state (Fe$^{4+}$) are known where Fe is surrounded by six oxygen atoms.\cite{8} BaFeO$_{3}$ (BFO) is one of the examples of the perovskite oxides with iron in valency +4 state.
In cubic crystal of BFO, ferromagnetism has been observed in the recent experiments.\cite{19} Ferromagnetism is found in pseudocubic BFO on SrTiO${_3}$(STO) films.\cite{20} There are also experimental reports on the successful growth of cubic BFO on STO.\cite{21,25,26}
Callender $et$ $al$.,\cite{26} have epitaxially grown  cubic BFO on STO and reported week ferromagnetism with transition temperature 235 K. Fully oxidized single crystal of BFO thin film also shows large saturation magnetization, in plane 3.2 $\mu_B$/formula unit (f.u) and out of plane 2.7 $\mu_B$/f.u. \cite{27} 
The reported lattice constant and saturation magnetization of BFO in thin films is quite close to bulk BFO(a = 3.97 $\rm \AA$) with no observed helical magnetic structure. \cite{27}
The absence of helical magnetic order in thin film might be due to small energy barrier between A-type helical magnetic order and ferromagnetic (FM) phases.\cite{28}
Very recently, we also found distortion-induced FM to antiferromagnetic (AFM) and ferrimagnetic transition in cubic BFO.\cite{p1} 

Tetragonal BFO is also of great interest as it may be a multiferroic phase of BFO,\cite{24} and Taketani $et$ $al$., \cite{21} have  reported the epitaxially grown BFO thin films on STO (100) substrate with a tetragonal crystal structure.
Hexagonal BaFeO$_{3-\delta}$ is expected to be the most stable phase although various polymorphs have been observed with oxygen deficit BFO.\cite{15,15A,15B,15C} Bulk hexagonal BaFeO$_{3-\delta}$ also exhibits an AFM to FM transition at 160 K.\cite{G2}
Using density functional theory (DFT), we found that strain and correlation also play a significant role in the magnetic and electronic properties of orthorrhombic BFO.\cite{34}
BFO can be alloyed with other magnetic perovskites such as BiFeO$_3$ to yield good multiferroic properties.

The well studied ferroelectric BaTiO$_3$ has different crystal structures, and shows different crystallographic behavior at different temperature and pressure.\cite{29} Similarly, the multiferroic BiFeO$_3$ also has different crystal structures.\cite{BiFeO3,a,b,c,d,e} However, there is no comprehensive theoretical calculations on less studied perovskite BaFeO$_3$  to investigate the true ground state structure of stoichiometric BFO with the help of DFT. Hence, we use DFT to investigate the equilibrium structure of BaFeO$_3$.

\section{Computatiioinal Method}
Calculations based on DFT are performed with plane-wave and pseudopotential method as implemented in the Quantum Espresso package.\cite{30} The exchange correlation effects are treated within the local spin density approximation (LSDA). The on-site Coulomb potential $U$(=5.0eV) \cite{31} has also been added in LSDA to perform LSDA+$U$ calculations to correctly describe the electronic structure of BFO in different crystallographic phases. The ultrasoft
pseudopotentials are used to describe the core-valence interactions. The valence wave functions and the electron density are described by plane-wave basis sets with kinetic energy cutoffs of 30 Ry with $12 \times 12 \times 12$ Monckhorst-Pack grid. All the computational parameters are fully converged. We studied BFO in four different crystal structures $i.e$, cubic, tetragonal, orthorhombic and rhombohedral with space groups \textit{Pm3m, P4mm, Amm2, R3m}, \cite{29} respectivily.

\section{Results and Discussions}
\begin{figure}
\subfigure[]{\includegraphics [width=0.3\textwidth]{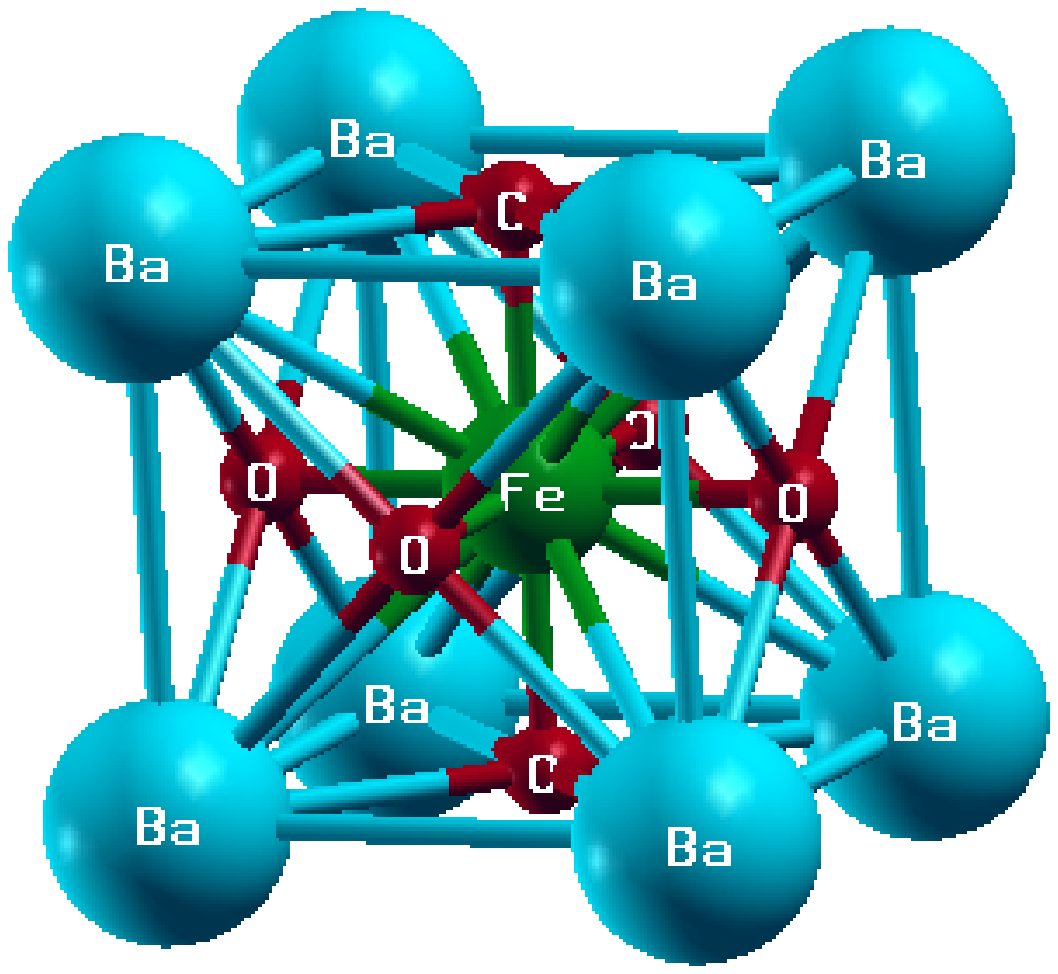}}
\subfigure[]{\includegraphics [width=0.3\textwidth]{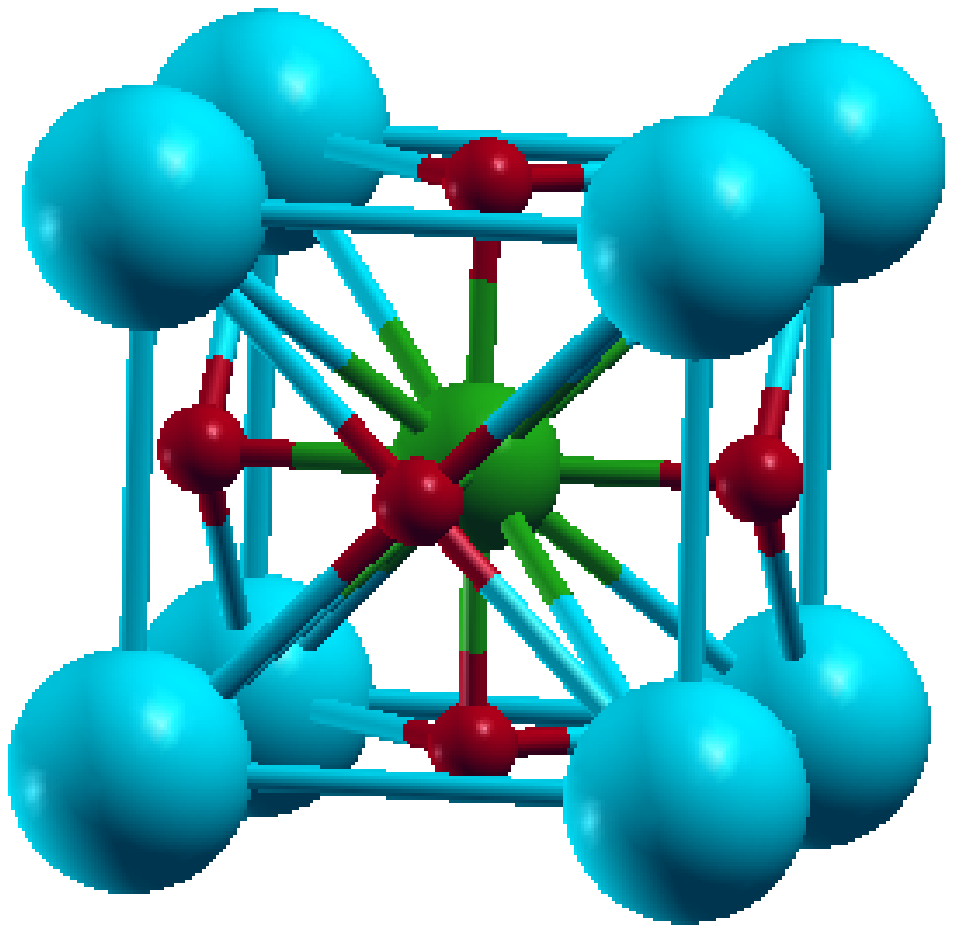}}
\subfigure[]{\includegraphics [width=0.3\textwidth]{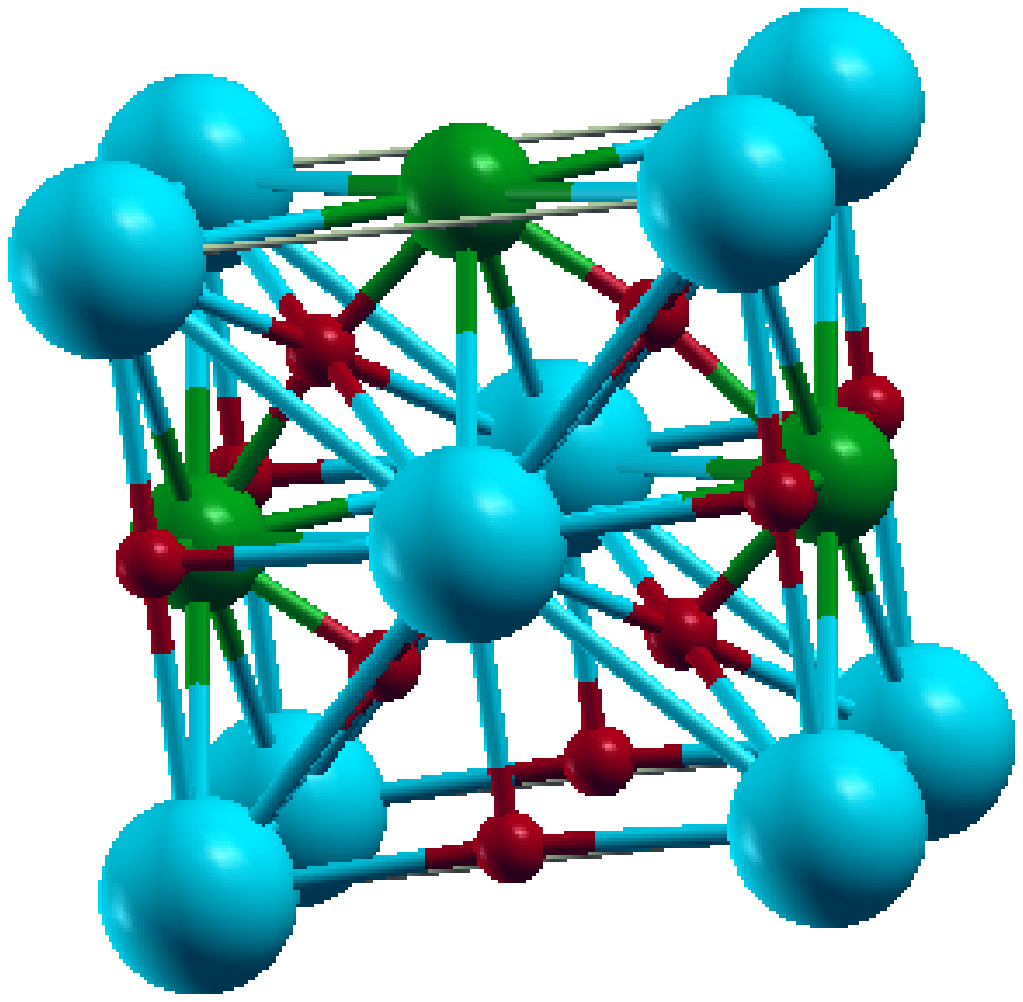}}
\subfigure[]{\includegraphics [width=0.3\textwidth]{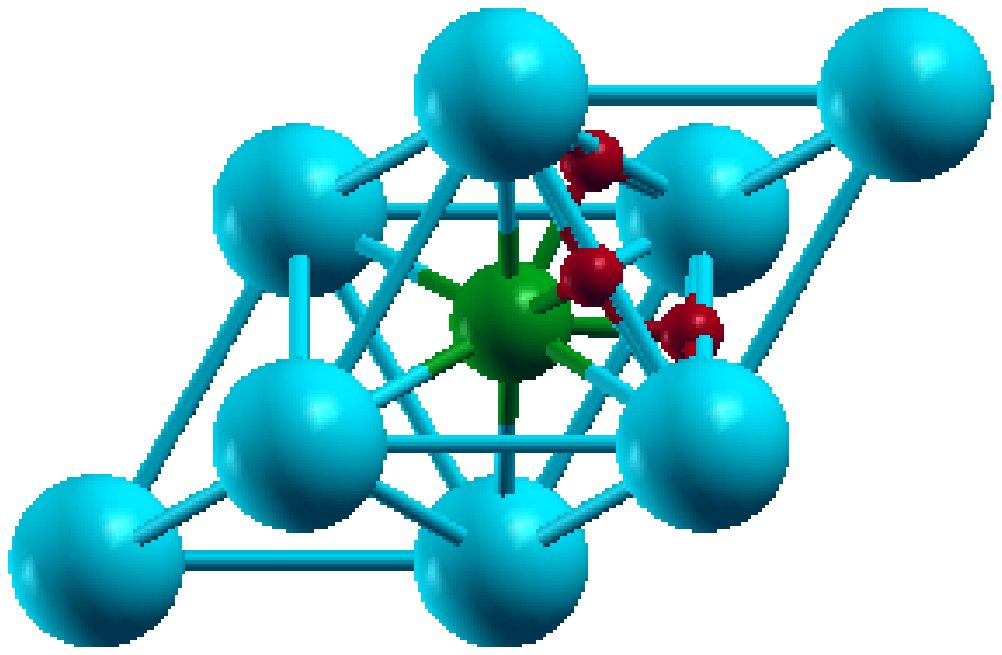}}
\caption{(Color online)Crystal structures of cubic(a), tetragonal(b), orthorhombic(c) and rhombohedral(d) BFO. All the structures have the same labelling of Ba, Fe, and O as in (a).}
\label{crys_str}
\end{figure}

\begin{figure}
\includegraphics [width=0.4\textwidth]{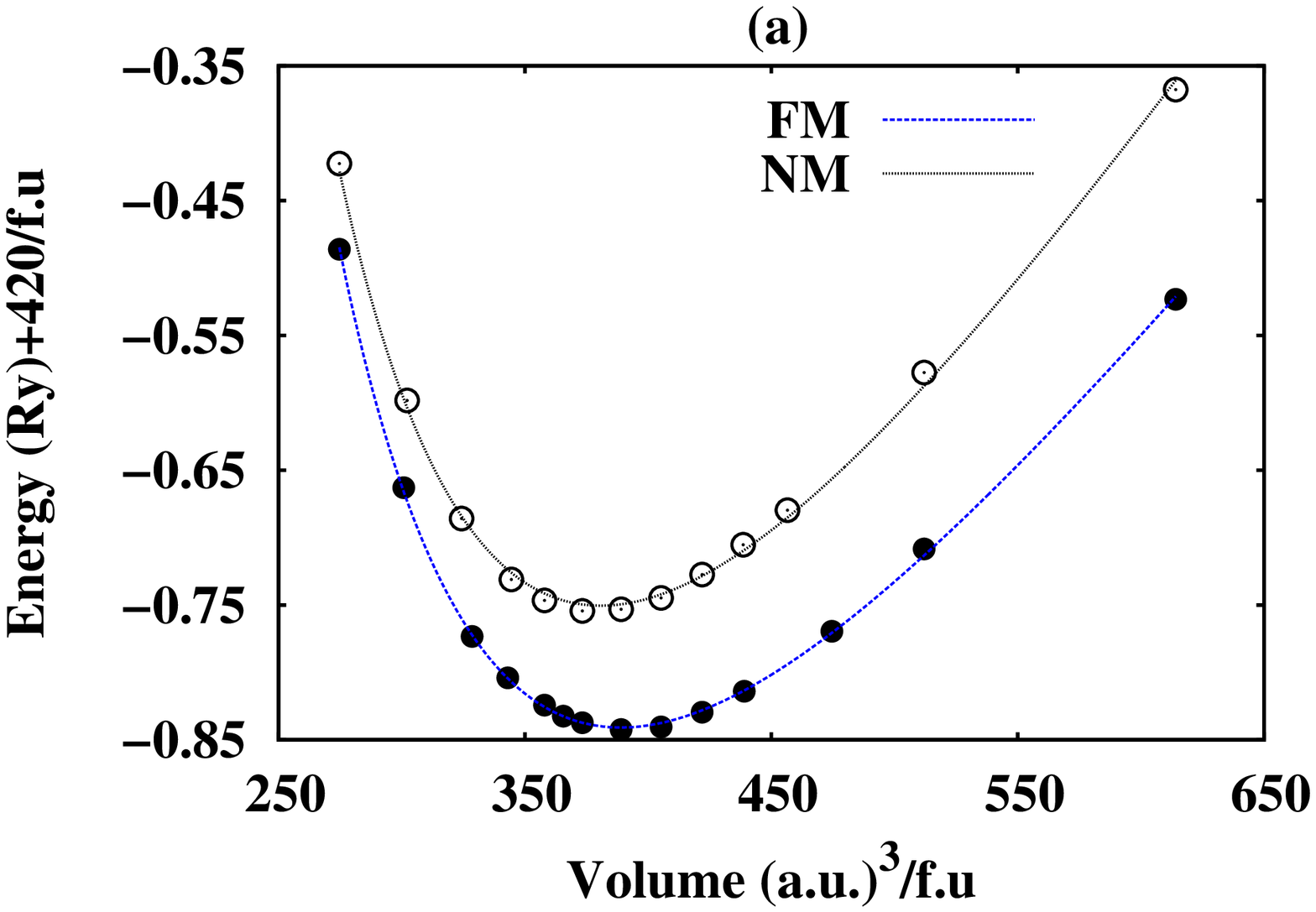}
\includegraphics [width=0.4\textwidth]{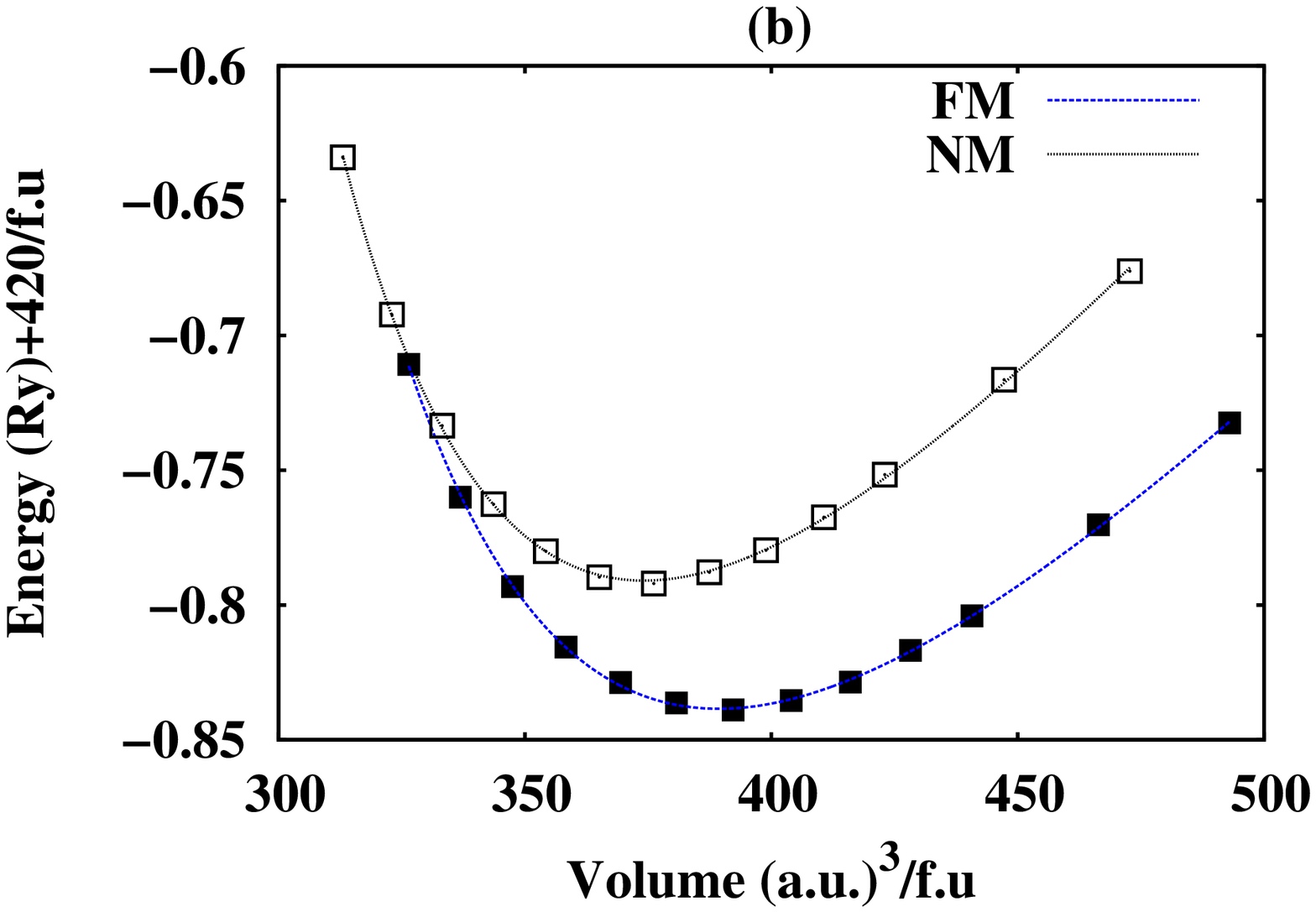}
\includegraphics [width=0.4\textwidth]{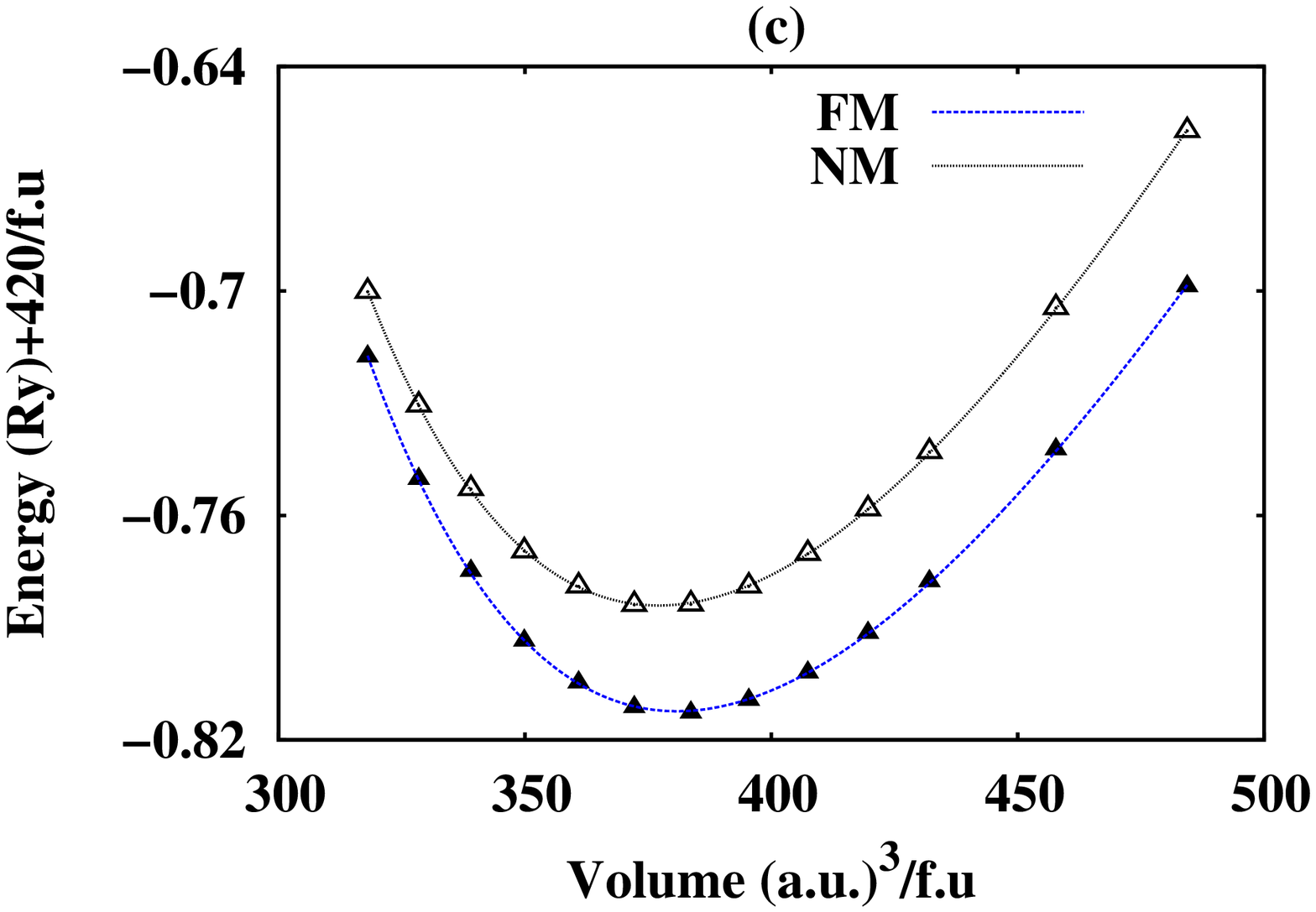}
\includegraphics [width=0.4\textwidth]{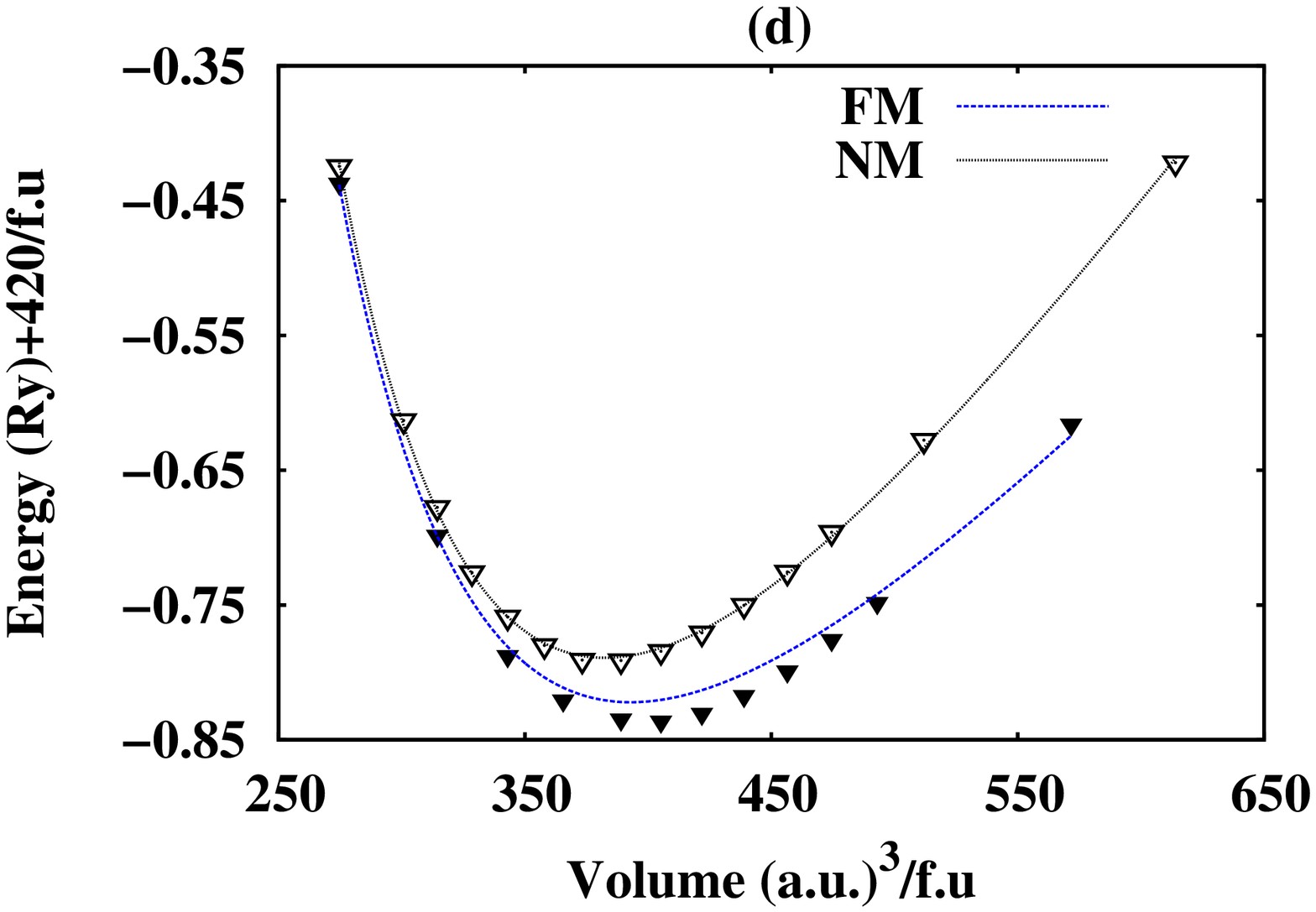}
\caption{(Color online) The LSDA calculated total energy(Ry) vs volume(a.u.$^3$) curves for cubic(a), tetragonal(b), orthorhombic(c) and rhombohedral(d) BFO. Filled (empty) symbols show FM(NM) states of BFO.}
\label{crys_vol}
\end{figure}

\begin{figure}
\includegraphics [width=0.4\textwidth]{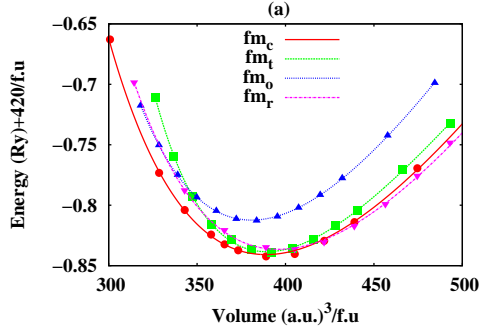}
\caption{(Color online) The LSDA calculated total energy(Ry) vs volume(a.u.$^3$) curves in FM states of cubic(fm$_c$), tetragonal(fm$_t$), orthorhombic(fm$_o$) and rhombohedral(fm$_r$) BFO.}
\label{vol_fit}
\end{figure}

	For different crystal structures (phases) of a material, it is very essential to optimize the lattice constants (volume) either using DFT or Molecular Dynamics. We used DFT and studied four different crystal structures of BFO $i.e$, cubic, tetragonal, orthorhombic, and rhombohedral which are shown in Fig.~\ref{crys_str} and their space groups are mentioned in Table~\ref{tab1}. To optimized the lattice volume, we carried out DFT calculations in ferromagnetic(FM) and non-magnetic(NM) states and then fitted the data using Birch Murnaghan equation of state (EOS) \cite{3} (shown in Fig.~\ref{crys_vol}), which enables us to estimate the equilibrium volume (lattice constant). These plots show that the FM state is more stable than the NM state in  all crystallographic phases of BFO. The optimized volumes and lattice parameters in the FM states are summarized in Table~\ref{tab1}. For comparison purpose, the combined energy volume (EV) curves in the FM states of all the studied structures are also shown in Fig.\ref{vol_fit}. From Fig.\ref{vol_fit} and Table~\ref{tab1}, it is inferred that the FM cubic BFO is the equilibrium (stable) structure of BFO among all the studied structures. This work is also supported by recently work on cubic BFO.\cite{32}		 
After confirming the stability of FM state w.r.t NM state, further calculations were performed to check the stability of FM w.r.t AFM state, in the (001) direction. We used the relation $\bigtriangleup E = E_{AFM} - E_{FM}$ [where $E_{AFM}(E_{FM}$) is the total energy in AFM(FM) state] to address the magnetic stability, and found that the FM state is more stable than the AFM state as shown in Table~\ref{tab1}. We see that cubic BFO has the largest $\bigtriangleup E $ which indicates possible room temperature ferromagnetism which is in agreement with the recent experimental work on cubic BFO.\cite{21,25,26}  

Once it is confirmed that the FM state is more stable than the NM and AFM states of BFO,  we calculated the formation energies of BFO in different structural phases.  The calculated formation energy in each phase confirms the lowest ground state energy. The formation energy is calculated by using the following formula, 
	
\begin{equation}
\bigtriangleup E_f =  E(\rm BaFeO_3) - [E(Ba) + E(Fe) + 3(\frac{1}{2})E(O_2)], 
\end{equation}	

where $E(\rm BaFeO_3)$ is the total energy in any studied crystallographic phase ($e.g.$ cubic) and $E$(Ba), $E$(Fe), and $E$(O$_2$) are the energies of BCC Ba, BCC Fe and oxygen molecule, O$_2$, respectivily.
The formation energy of each system is shown in the Table~\ref{tab1}. The formation energy of cubic phase is minimum (-3.85 eV) as noted in the EV curve (Fig.\ref {vol_fit}) that cubic structure is the most stable structure of BFO. At the same volume (equilibrium volume of cubic BFO), the second most stable structure is tetragonal BFO (Fig.\ref {vol_fit}). The small energy difference between cubic BFO and tetragonal BFO clearly indicates  a possible phase transition to tetragonal BFO. Such a small energy difference can easily be recovered if the cubic BFO is grown as a thin film on a suitable substrate, e.g., SrTiO$_{3}$.\cite{21} Note that such a possible tetragonal BFO is supported by the experimental work.\cite{21} 
Furthermore, when we expand the cubic unit cell, it cuts rhombohedral phase at volume slightly larger than the equilibrium volume of cubic phase, so the third stable (metastable) phase is the rhombohedral phase consistent with the formation energy (Table~\ref{tab1}). Note that similar crystal stability  was also observed in BiFeO$_{3}$.\cite{BiFeO3}

\begin{figure}
\includegraphics [width=0.4\textwidth]{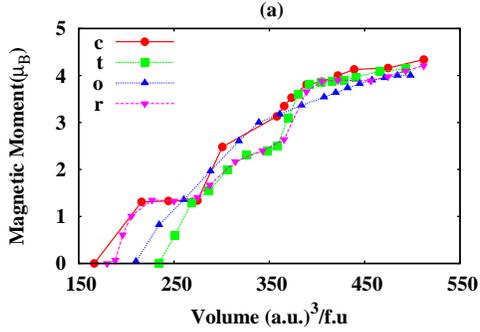}
\caption{(Color online) The LSDA calculated total magnetic moment/fu(in units of $\mu_B$) vs volume/fu (a.u.$^3$) for cubic(circle), tetragonal(square), orthorhombic(up triangle) and rhombohedral(down triangle) BFO.}
\label{mag}
\end{figure}

The calculated magnetic moment (MMs)/f.u  of BFO in different crystal structures are shown in Fig.\ref{mag}, and in Table~\ref{tab2} the values of total and local MMs of Fe and O atoms in each phases of BFO at equilibrium lattice volume are summarized. In Fig.\ref{mag}, for all the phases of BFO, the magnetic moment increases with increasing volume and at some particular volume (near equilibrium volume), it attains almost a constant value. The increase in MM with lattice constant usually happens due to the decrease in the overlap between the orbitals, $i.e.,$ Fe and O atoms. The comparison of total magnetic moments of all the structures (Table~\ref{tab2}) shows that the orthorhombic BFO has the lowest MM value due to its smaller volume/f.u  as shown in Table.~\ref{tab1} The total magnetic moments of cubic, tetragonal, and rhombohedral phases are approximately the same because their equilibrium volumes are slightly different from each other. The local MMs show that the major contribution to the total moment is coming from the Fe atoms.  Small induced MMs at O sites due to Fe-O hybridization can also be seen. In cubic and rhombohedral systems, the  Fe-O bond lengths are the same so the local moments are also the same for all the three oxygen atoms. In the remaining structures, due to different Fe-O bond length, oxygen atoms have different magnetic moments. These different local moments of oxygen show different crystal symmetry of BFO. In BFO, Fe has four unpaired electrons, all with high spin (spin up) state. This suggests that BFO should have total magnetic moment of 4$\mu_B$, but this is not the situation. The deviation from 4$\mu_B$ is due to the strong hybridization of Fe with oxygen atoms. However, LDSA+$U$ calculation shows 4$\mu_B$ which suggests that the strong Fe-O hybridization reduces the magnetic moment. Including the Columb-type repulsive interaction, i.e., $U$, decreases the  Fe-O hybridization that gives  an integer magnetic moment value in each phase of BFO. The non-integer (integer) values of the MMs in LSDA (LSDA$+U$) is also confirmed in the electronic structures which are discussed in the the following paragraphs.

\begin{figure}
\includegraphics [width=0.4\textwidth]{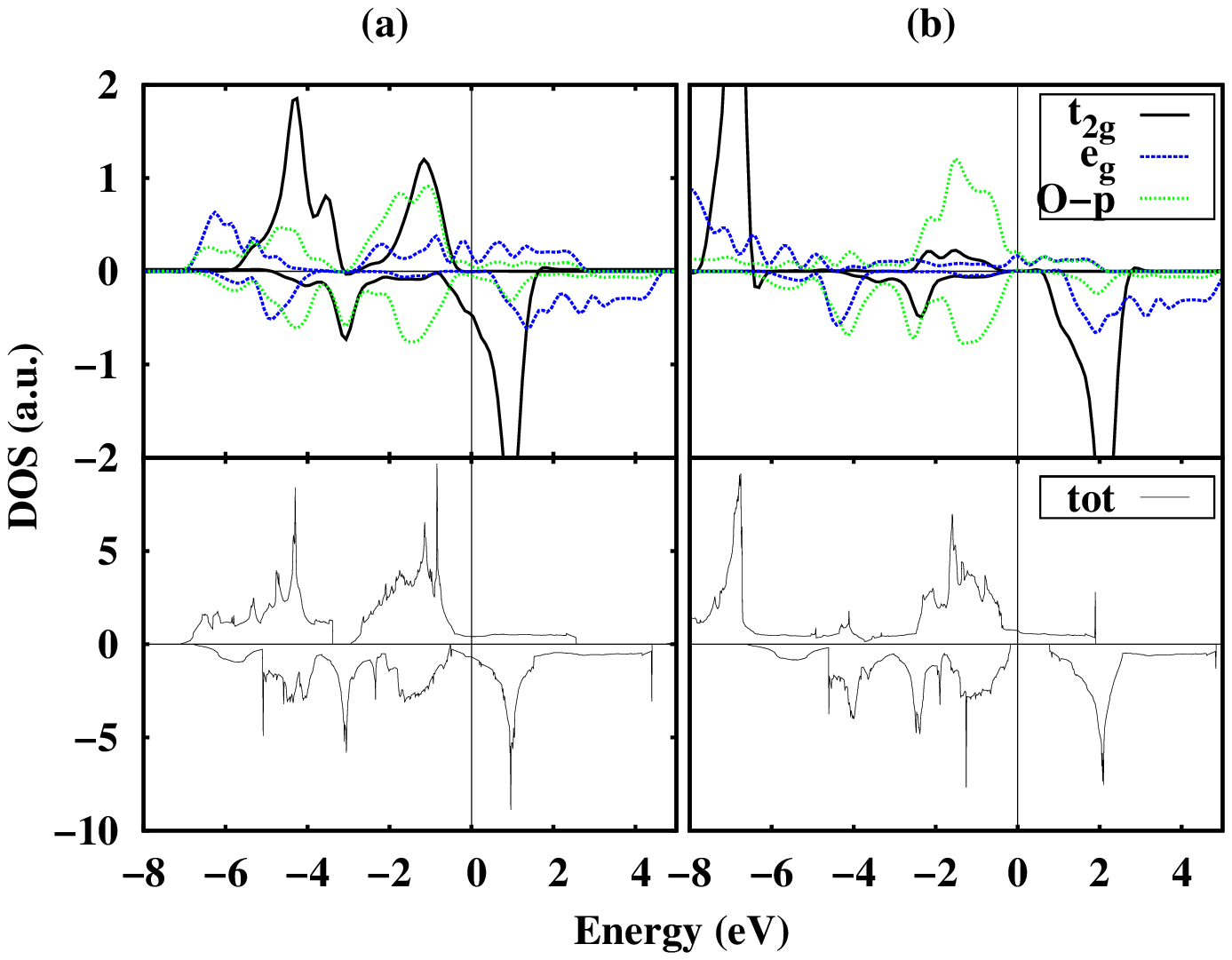} \hspace{1.5cm}
\vspace{1.5cm}
\includegraphics [width=0.4\textwidth]{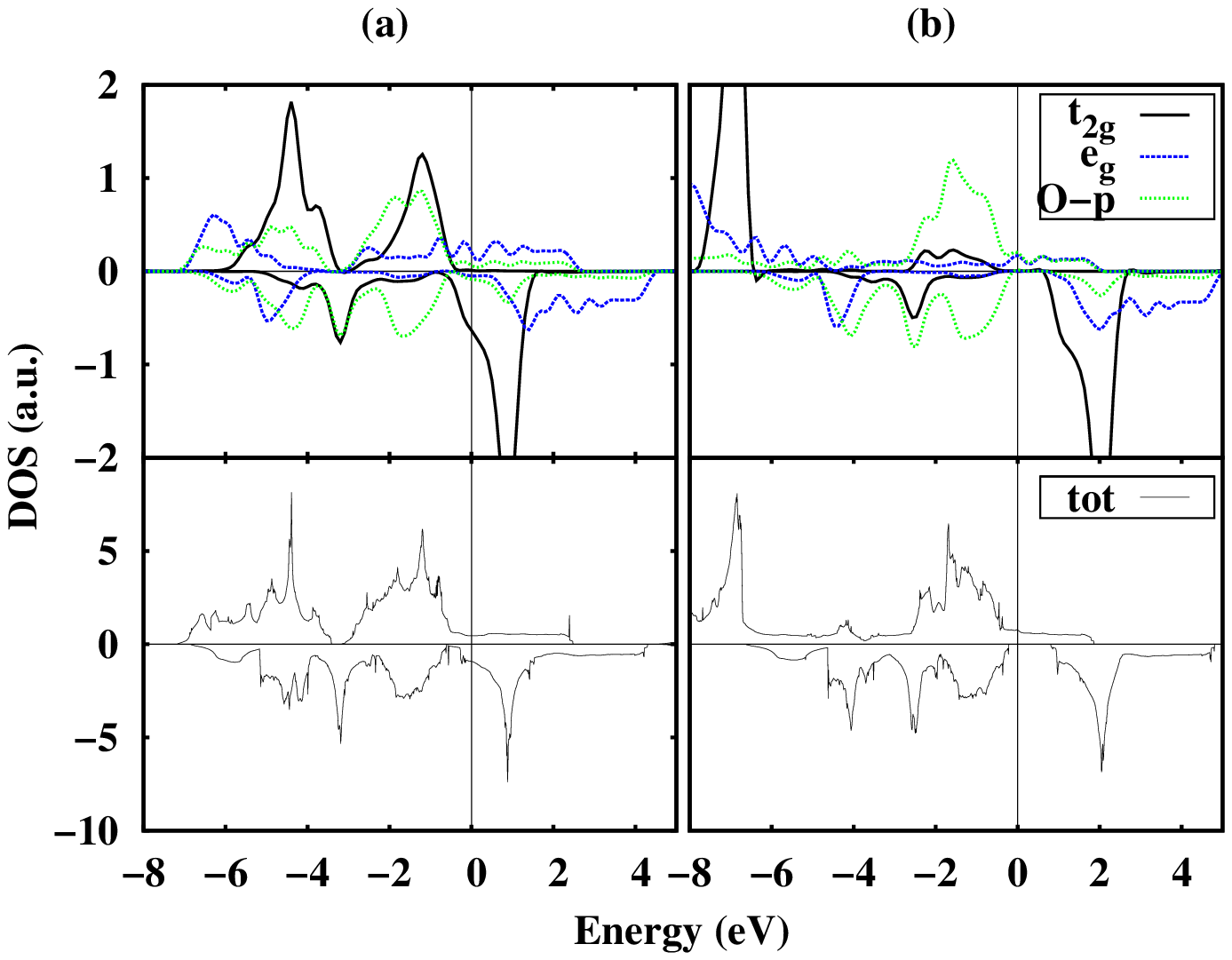}
\includegraphics [width=0.4\textwidth]{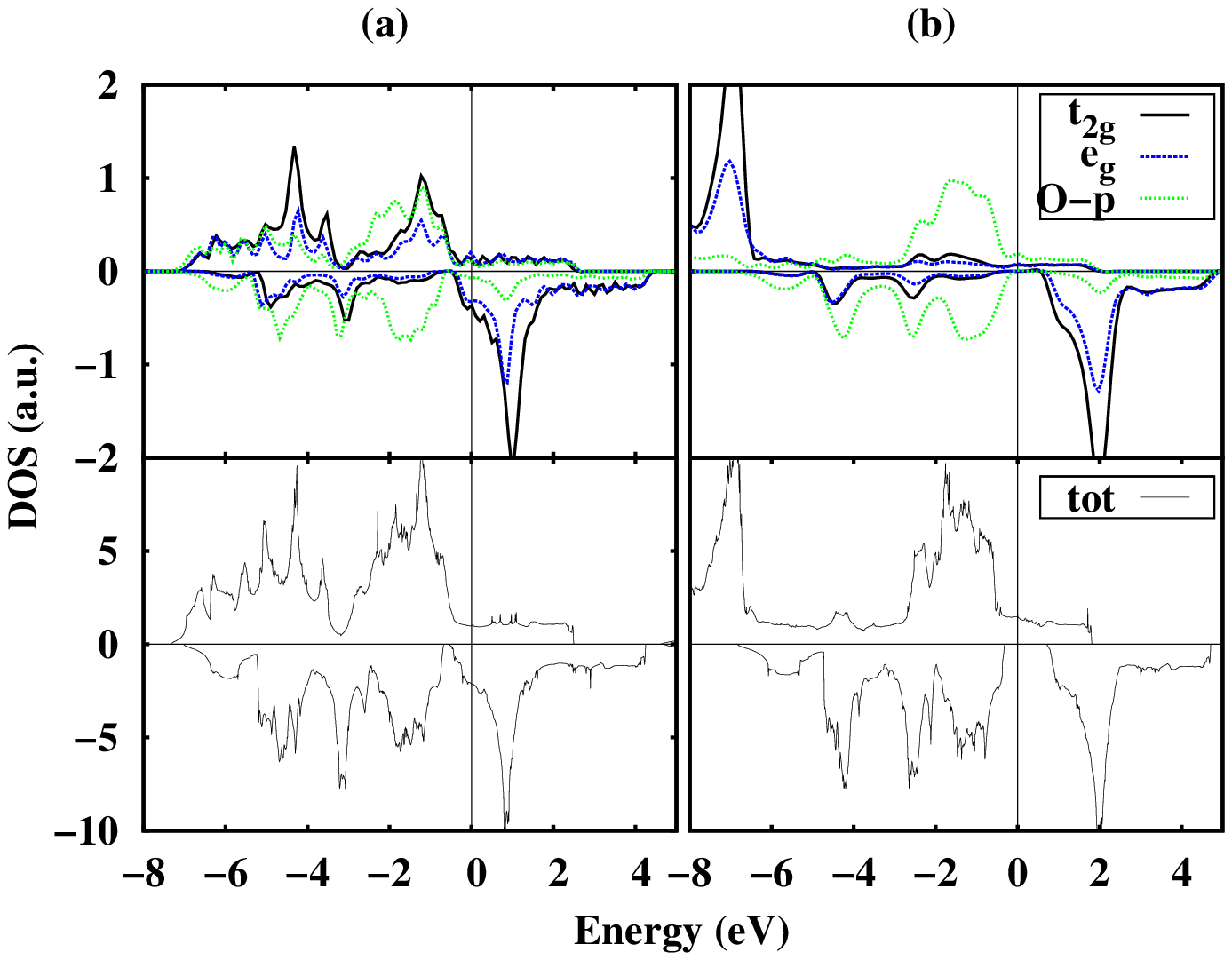} \hspace{1.5cm}
\vspace{1.5cm}
\includegraphics [width=0.4\textwidth]{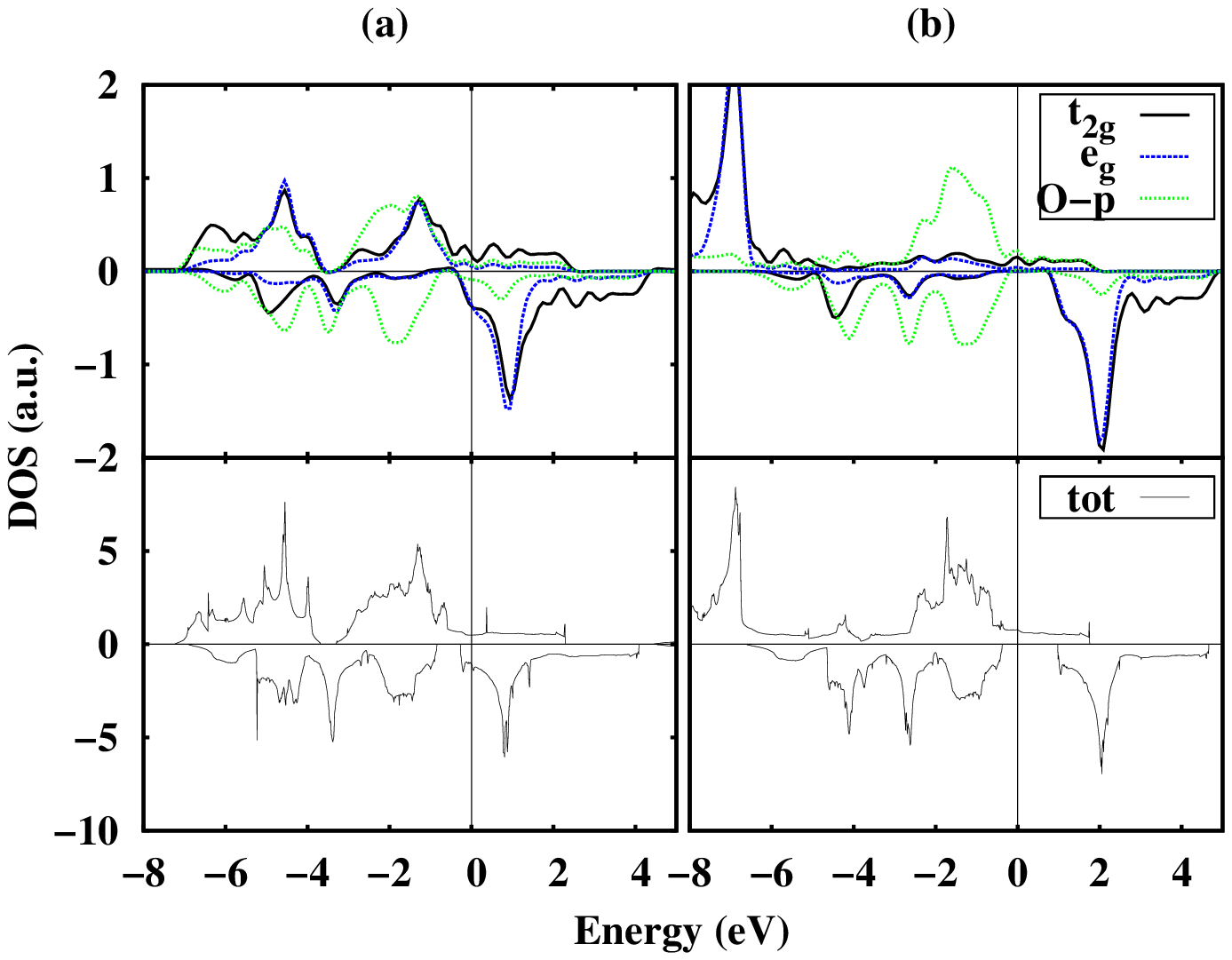}
\caption{(Color online) The calculated total density of states(DOS) and projected density of states(PDOS) in cubic(up left), tetragonal(up right), orthorhombic(down left) and rhombohedral(down right) BFO. The left(right) pannel shows the result of LSDA(LSDA+$U$). The Fermi energy is set at zero eV.}
\label{dos}
\end{figure}

The electronic properties are also investigated in each phase of BFO using both LSDA and LSAD$+U$ (see Fig.~\ref{dos}). We noticed that the electronic properties of BFO in all studied phases are showing almost similar behaviour and there is a transformation from metallic to half-metallic phase in each case by including $U$. The total density of states (DOS) and projected density of states (PDOS) of cubic BFO in LSDA show that cubic BFO is a metal consistent with the non integer value of the magnetic moment. The contribution to the total DOS separately from Fe and O atoms is also shown in the PDOS. The Fe $d$ orbital is further splitted into doubly degenerate $e_g$ and triply degenerate $t_{2g}$ states. From PDOS it can be seen that there is a strong hybridization between Fe-$e_g$ and O-$p$ states, while the Fe-$t_{2g}$ and O-$p$ states are weakly hybridized. The strong coupling between Fe-$e_g$ and O-$p$ states is due to direct overlapping of these orbitals. The $t_{2g}$ states have minor contributions in the majority spin states at the Fermi level. Such hybridization of Fe and O orbitals in cubic BFO generates metallicity.  However, including $U$ has quite different effect on the Fe-$d$ orbitals. The DOS and PDOS for LSDA+$U$ is also plotted in the same Figure~\ref{dos}(b). Due to the introduction of correlation energy through the Hubbard-$U$ term, a different behavior is appeared in the minority spin states, creating a band gap of approximately 0.95 eV at the Fermi level, giving a half metallic character to the cubic BFO. Such half metallic behavior in perovskites is very important from application point of view. The Hubbard-$U$ term has a small effect on Fe-$e_g$ bands as these states are strongly hybridized with the O-$p$ states while increasing the localization of $t_{2g}$ bands. This (Hubbard $U$ term) reduces the already small $t_{2g}$ electrons hybridization with O-$p$ states and shifting both the minority and majority $t_{2g}$ states further away from the Fermi level. The shifting is from -5 eV to -7 eV below the Fermi level in the spin up $t_{2g}$ states, and from 1 to 2 eV above the Fermi level in the spin down $t_{2g}$ states. Our results are in well agreement with the previous \textit{ab-initio} calculations.\cite{33}
 
The electronic structures of tetragonal, orthorhombic, and rhomohedral BFO are also analysed, and the electronic structures are almost similar to cubic BFO. The DOS and PDOS for tetragonal BFO is plotted in Fig.\ref{dos}, which shows the same behaviour as the cubic structure but the band gap in spin down for LSDA+$U$ is 1.02 eV which is larger than the cubic BFO band gap. Similarly, the orthorhombic BFO is shown in Fig.\ref{dos} and is showing the same results as for cubic and tetragonal except a prominent pseudo gap in LSDA in minority spin states just below the Fermi level which can also be shifted to Fermi level by applying external strain.\cite{34} However, in LSDA+$U$ calculations the band gap in the minority spin state is 1.03 eV. We also expect that rhombohedral BFO (Fig.\ref{dos}) may show half-metallic behavior under uniaxial strain, similar to orthorhombic BFO.\cite{34} The LSDA+$U$ calculated half-metallic band gap is 1.3 eV, which is larger than the other phases of BFO due to larger volume of orthorhombic BFO.

\section{Conclusion}
Density functional theory is used to predict the ground state crystal structure of BaFeO$_{3}$. Local spin density approximation (LSDA) was used for the exchange and correlation functional. Different crystal structures (cubic, tetragonal, orthorhombic, and rhombohedral) of BFO are considered. LSDA calculations showed that cubic BFO has the lowest formation enthalpy among the studied crystal structures. It is also observed that FM states of BFO are more stable as compared with NM and AFM states. The electronic structures within LSDA showed that all phases of BFO are metallic. To correctly describe the electronic band structures, further calculations were carried out using the  LSDA+$U$ approach. The  LSDA+$U$ calculated band structures of BFO are half-metallic. The LSDA+$U$ approach showed that  adding Hubbard like potential $U$ deceased  the hybridizations between the Fe $d$ and O $p$ orbitals, and such reduced hybridization resulted half-metallicity in BFO. The calculated magnetic moments showed integer (non-integer) values in 
LSDA+$U$ (LSDA)calculations.
\section{Acknowledgment}
	We acknowledge National Centre for Physics (NCP) Islamabad, Pakistan for providing computing facilities.

{}
	 
\newpage

\begin{table}
\centering
\caption{The LSDA calculated total volume per formula unit (vol/f.u), lattice parameters $(a, b, c)$, angles $(\alpha, \beta, \gamma)$, formation enthalpy ($\bigtriangleup$E$_f$), energy difference between AFM and FM states($\bigtriangleup E$ $E_{AFM}-E_{FM}$) and space groups of cubic, tetragonal, orthorhombic and rhombohedral crystal structures. The volume (lattice parameter) is in units of a.u.$^3$ (a.u.). The $\bigtriangleup E_f$ and $\bigtriangleup E$  are in units of eV}.

\begin{tabular}{c|cccccccccc}
\hline \hline
System & Vol/f.u & $a$ & $b$ & $c$ & $\alpha$ & $\beta$ & $\gamma$ & $\bigtriangleup E_f$ &  $\bigtriangleup E $  &  space group \\ \hline \hline
Cubic & 388.87 & 7.30 &  7.30 &  7.30 & 90$^o$ & 90$^o$ & 90$^o$ & -3.85 &   0.25 & Pm3m \\ 
Tetragonal & 389.55 & 7.28 &  7.28 &  7.35 & 90$^o$ & 90$^o$ & 90$^o$ & -3.84 &   0.07 & P4mm \\ 
Orthorhombic & 380.77 & 7.20 & 10.25 & 10.29 & 90$^o$ & 90$^o$ & 90$^o$ & -3.80 &   0.13 & Amm2 \\ 
Rhombohedral & 392.93 & 7.32 &  7.32 &  7.32 & 90$^o$ & 90$^o$ & $<$ 90$^o$  & -3.83 &   0.03 & R3m \\ \hline 
\end{tabular}
\label{tab1}
\end{table}

\begin{table}[:h]
\centering
\caption{The LSDA calculated Total Magnetic Moment of the system per f.u. and the local magnetic moment of Fe, O$_1$, O$_2$ and O$_3$. Total (local) moment is in units of $\mu_B$.}

\begin{tabular}{c|ccccc}
\hline \hline
System & Total Moment of System/f.u & Fe & O$_1$ & O$_2$ & O$_3$ \\ \hline \hline
Cubic & 3.46 & 2.83 & 0.20 & 0.20 & 0.20 \\ 
Tetragonal & 3.44 & 2.81 & 0.23 & 0.19 & 0.19 \\ 
Orthorhombic & 3.34 & 2.74 & 0.19 & 0.20 & 0.20 \\ 
Rhombohedral & 3.42 & 2.80 & 0.21 & 0.21 & 0.21 \\ 
\hline

\end{tabular}
\label{tab2}
\end{table}

\clearpage

\end{document}